\documentclass{aa}

\usepackage{graphicx}
\usepackage{txfonts}
\usepackage{xcolor}
\usepackage{xspace}
\usepackage{hyperref}
\usepackage{comment}
\hypersetup{
     colorlinks   = true,
     citecolor    = blue
}

\def \inte {{\em INTEGRAL}}
\def \ibis  {{\em IBIS/ISGRI}}
\def \swift {{\em Swift}}

\def \maxi{{\em MAXI J0911$-$655}}
\def \igr{{\em IGR J16597$-$3704}}

\def \swiftxrt{{\em Swift/XRT}}
\def \nustar{{\em NuSTAR}}

\def \Msun{{M$_{\odot}$}}

\begin{document}

\title{Discovery of 105 Hz coherent pulsations in the ultracompact binary \igr{}}

   \author{A. Sanna\inst{1},
   	  A. Bahramian\inst{2},
	  E. Bozzo\inst{3},
	  C. Heinke\inst{4},
	  D. Altamirano\inst{5},
	  R. Wijnands\inst{6},
	  N. Degenaar\inst{6},
	  T. Maccarone\inst{7},
	  A. Riggio\inst{1},
	  T. Di Salvo\inst{8},
	  R. Iaria\inst{8},
	  M. Burgay\inst{9},
	  A. Possenti\inst{9},
	  C. Ferrigno\inst{3},
          A. Papitto\inst{10},
          G. R. Sivakoff\inst{4},
          N. D'Amico\inst{1,9}
         L. Burderi\inst{1}
          }

   \institute{Dipartimento di Fisica, Universit\`a degli Studi di Cagliari, SP Monserrato-Sestu km 0.7, 09042 Monserrato, Italy\\
   		\email{andrea.sanna@dsf.unica.it}
   	\and
	       Department of Physics and Astronomy, Michigan State University, East Lansing, MI, USA
	 \and      
	       ISDC, Department of Astronomy, University of Geneva, Chemin d'\'Ecogia 16, CH-1290 Versoix, Switzerland
         \and
              University of Alberta, Physics Dept., CCIS 4-181, Edmonton, AB T6G 2E1
         \and
               Physics and Astronomy, University of Southampton, Southampton, Hampshire SO17 1BJ, UK
         \and
               Anton Pannekoek Institute for Astronomy, University of Amsterdam, Science Park 904, 1098 XH, Amsterdam, the Netherlands
          \and
              Department of Physics, Box 41051, Science Building, Texas Tech University, Lubbock TX, 79409-1051, USA 
          \and
               Universit\`a degli Studi di Palermo, Dipartimento di Fisica e Chimica, via Archirafi 36, 90123 Palermo, Italy  
         \and
                 INAF, Osservatorio Astronomico di Cagliari, Via della Scienza 5, I-09047 Selargius (CA), Italy
         \and        
                 INAF, Osservatorio Astronomico di Roma, Via di Frascati 33, I-00044, Monteporzio Catone (Roma), Italy        
             }

   \date{Received -; accepted }

  \abstract
   {We report the discovery of X-ray pulsations at 105.2 Hz (9.5 ms) from the transient X-ray binary \igr{} 
   using \nustar{} and \swift{}. The source was discovered by \inte{} in the globular cluster NGC 6256 at a distance of 9.1 kpc. The X-ray pulsations show a clear Doppler modulation implying an 
orbital period of $\sim$46 minutes and a projected semi-major axis of $\sim5$ lt-ms, which makes \igr{} an ultracompact X-ray binary system. 
We estimated a minimum companion mass of $6.5\times10^{-3}$ M$_{\odot}$, assuming a neutron star mass of 1.4 \Msun, 
and an inclination angle of $<$ 75$^{\circ}$ (suggested by the  absence of eclipses or dips in its light-curve). The broad-band energy spectrum 
of the source is well described by a disk blackbody component (kT $\sim$1.4 keV) plus a comptonised power-law with photon index $\sim$2.3 and an 
electron temperature of $\sim$30 keV. Radio pulsations from the source were searched for with the Parkes observatory and not detected.}


\titlerunning{new AXMP \igr{}}
\authorrunning{Sanna et al.}

   \maketitle

\section{Introduction}

So far, 19 accreting millisecond X-ray pulsars (AMXP) were discovered in low mass X-ray binaries, with spin periods ranging from  
1.7 ms to 6.1 ms \citep[e.g.,][]{Burderi13, Patruno2017a}. The short 
periods of these sources are the result of the long-lasting mass transfer from the companion star via Roche-lobe overflow   
\citep{Alpar82}, a scenario that was observationally confirmed by the discovery of transitional 
binary pulsars \citep[see][and references therein]{Papitto2013b}.  
Almost a third of the total sample of AMXPs is classified as \emph{ultracompact} binary systems, showing extremely short orbital periods 
\citep[P$_{{\rm orb}} < 1$ hrs;][]{Galloway02, Markwardt02, Kirsch04, Krimm07, Altamirano2010a, Sanna2017a}. The rest of the sample have  
P$_{{\rm orb}} < 12$ hrs, except for Aql X$-$1 \citep[P$_{{\rm orb}} \sim 19$ hrs;][]{Welsh2000a}. 
The distribution of short orbital periods suggests that typical AMXP companion star have masses below 0.2~M$_{\odot}$ \citep{Patruno12b}.

We report here on the discovery of millisecond X-ray pulsations from \igr{}, a transient source discovered by \inte{} 
on 2017 October 21 \citep{Bozzo2017a} and localised within the globular cluster NCG 6256 at 9.1~kpc \citep{Bozzo2017b, Valenti2007a}. 
The best known position of the source is at RA = $16^h 59^m 32.9007^s \pm 0.0021^s$, DEC=$-37^{\circ} 07\arcmin 14.22\arcsec \pm 0.18\arcsec$  
\citep{Tetarenko2017a}. 

\section[]{Observations and data reduction}

\subsection{\nustar{}}
\nustar{} \citep[][]{Harrison2013a} observed \igr{} (Obs.ID. 90301324001) on 2017 October 26 at 12:16 UTC for an elapsed time of $\sim84$ ks, corresponding to a total exposure of $\sim42$ ks. 
We performed standard screening and filtering of the events using the \nustar{} 
data analysis software (\textsc{nustardas}) version 1.5.1. We extracted source events from the two focal plane modules (FMPA and FMPB) within a circular 
region of radius 100$''$ for the light curve and 60$''$ for the spectrum, centered on the source position. Similar regions, but centered away from the source, 
were used to extract the background products.
Response files were generated using the \textsc{nuproducts} pipeline and the photon arrival times were barycenter-corrected by using 
the \textsc{barycorr} tool. No Type-I bursts were recorded during the observation. 

\subsection{\inte{}}
\label{sec:integral}

We considered all the public available \inte\ science windows for IBIS/ISGRI 
\citep[20-100~keV,][]{lebrun03,ubertini03} that were performed in the 
direction of \igr{} during the outburst (i.e., revolutions 1876 and 1878, covering from 2017 October 
21 at 04:40 to October 27 at 14:51, UTC). During this period, the source was not detected by JEM-X \citep{lund03} due to the large off-axis 
position \citep{Bozzo2017a}.
All the \inte{} public data were analysed using version 10.2 of the OSA 
software \citep{courvoisier03}. \igr{} was detected in the total IBIS/ISGRI 20-40~keV mosaic at 10~$\sigma$. We extracted a single ISGRI 
spectrum using the entire exposure time available (53~ks).  

\subsection{\swift{}}
\label{sec:swift}
We used one \swiftxrt{} \citep{Gehrels04} observation of \igr{} taken on 2017 October 25 at 07:51 (UTC) 
in WT mode (1 ks). We processed the XRT data with \textsc{xrtpipeline} version 0.13.4, extracted source and background spectra using \textsc{xselect}, and produced 
an  ancillary response file for the spectrum with \textsc{xrtmkarf}. We used circular extraction regions with radii of 20 pixels ($\sim 47''$) for both the source 
and background, considering only grade 0 events to avoid low-energy calibration issues. 

\subsection{Radio Observations with Parkes}
\label{sec:parkes}
\igr{} was observed for 2.9 hrs on 2017 November 3 at the Parkes radio telescope with the aim of searching for radio pulsations. The data were collected using the BPSR backend over a bandwidth of 400 MHz (reduced to 315 after interference removal) centered at 1382 MHz and split into 1024 frequency channels. The signal was 2-bit sampled every 64 $\mu$s.  Data were folded using the X-ray ephemeris of Tab.~\ref{tab:solution} and searched over dispersion measures DM $<$ 700 pc/cm$^3$ \citep[the nominal DM expected for NGC 6256 ranges from 250 to 450 pc/cm$^3$;][]{Taylor1993a, Cordes2002a, Schnitzeler2012a, Yao2017a}. Data were also folded with different trial values in T$_{{\rm NOD}}$ while the uncertainty on the other parameters was neglected as it would affect the width of the folded profile by < 10\%.
No radio pulsation were found down to a flux density of 0.05 mJy.

\section{Data analysis}

\subsection{Timing analysis}
\label{sec:ta2017}

To search for coherent signals we extracted a power density spectrum (PDS) from the \nustar{} observation by averaging together 9 power 
spectra produced over 4096 seconds of data (Fig.~\ref{fig:pds}). The PDS shows a highly significant (>100$\sigma$) broad double-peaked 
signal with central frequency $\sim 105.175$ Hz and width $\sim1.5\times 10^{-3}$ Hz. Moreover, two prominent spikes at $\sim210$ Hz and $\sim315$ Hz (harmonically related with the strongest peak) are also clearly visible. To produce an orbital solution for the neutron star (NS), we created power density spectra every 512 s and inspected them for significant features in the $1.5\times 10^{-3}$ Hz interval around the central frequency value reported above. We modelled the observed spin frequency variation of the signal with respect to the central spin frequency as the binary orbital Doppler shift: 
\begin{figure}
\centering
\includegraphics[width=0.45\textwidth]{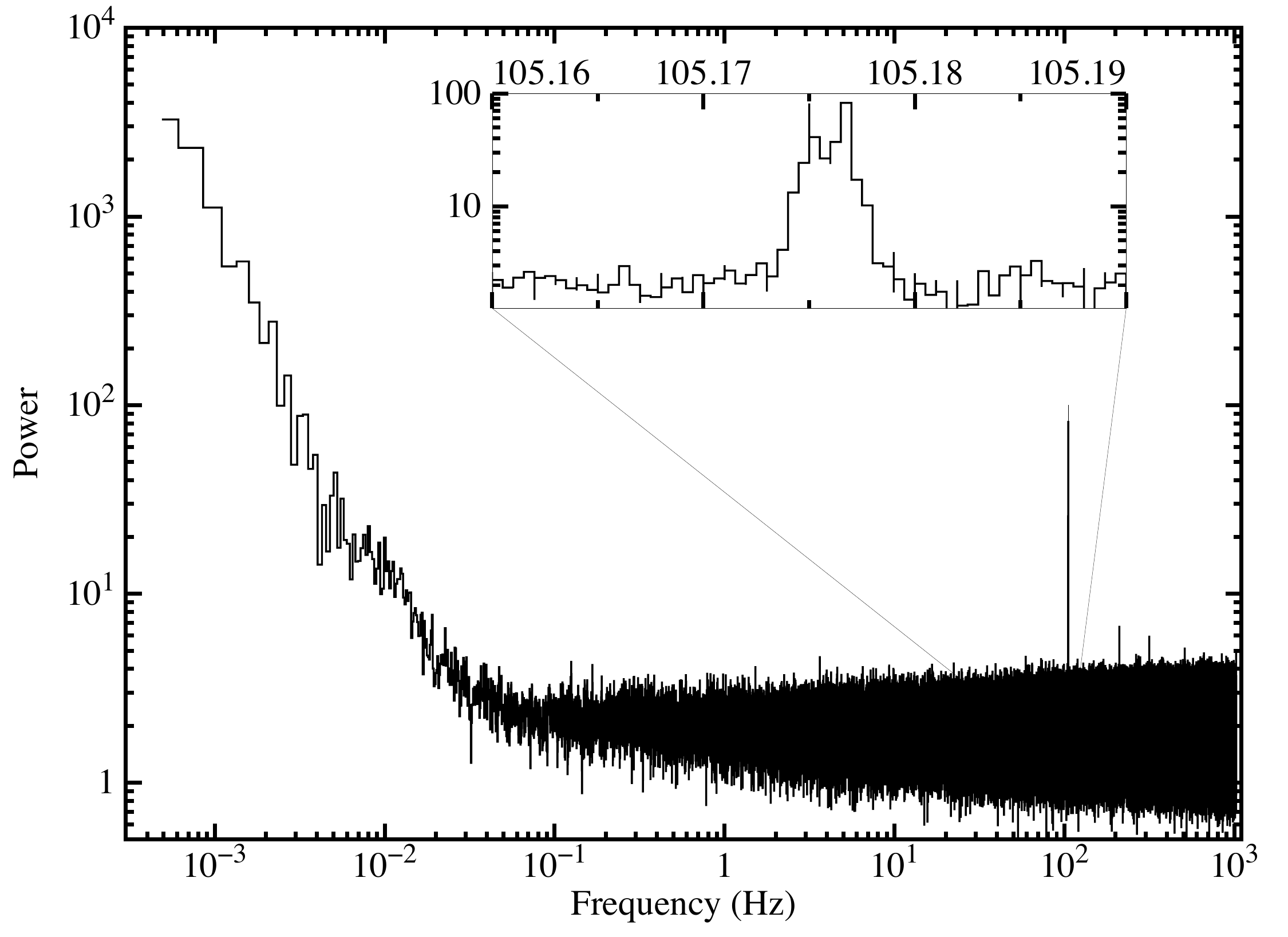}
\caption{Leahy normalised \citep{Leahy1983a} PDS, produced by averaging 4096 s-long segments of \nustar{} data. The 
fundamental frequency ($\sim$105 Hz), as well as the second and third harmonic, are clearly visible in the power spectrum. The 
inset shows a zoom of the double-peaked profile of the fundamental frequency.}
\label{fig:pds}
\end{figure} 
\begin{equation}
\Delta \nu = \Delta \nu_0 - \frac{2 \pi \, \nu_0 \, x}  {P_{\rm orb}}  \cos\left(\frac{2 \pi \,(t - T_{\rm NOD})} { P_{\rm orb}}\right)
\end{equation}
where $\nu_0$ is the spin frequency, $x$ is the projected semi-major axis of the NS orbit in light seconds, $P_{\rm orb}$ is the orbital period, 
and $T_{\rm NOD}$ is the time of passage through the ascending node. We found: $P_{\rm orb}=2751(2)$ s,  
$x = 0.0052(2)$ lt-s, $T_{\rm NOD}=58052.4898(5)$ (MJD), and $\nu_0=105.17582(4)$ Hz.

We then corrected the photon time of arrivals for the binary motion by applying the orbital ephemeris previously reported through the recursive formula  
\begin{eqnarray}
\label{eq:barygen} 
t + x\,\sin\Big(\frac{2\pi}{P_{\rm orb}} \,(t-T_{\rm NOD})\Big) = t_{\rm arr}
\end{eqnarray}
where $t$ is the photon emission time, $t_{\rm arr}$ is the photon arrival time to the Solar System barycentre, and the second term on 
the left side of Eq.~\ref{eq:barygen} represents the projection along the line of sight of the distance between the NS and the barycenter 
of the binary system in light seconds assuming almost circular orbits (eccentricity $e \ll 1$).
\noindent
We epoch-folded segments of $\sim$250 seconds in 16 phase bins at the frequency  $\nu_0=105.17582(4)$ Hz obtained above. 
Each pulse profile was modelled with a sinusoid of unitary period, to determine the corresponding amplitude and the fractional part of the epoch-folded phase residual. 
We selected only profiles with the ratio of the amplitude of the sinusoid to its 1~$\sigma$ uncertainty larger than three. 
The second and third harmonic components were only significantly detected in 
less than 10\% of the total number of intervals.

\begin{figure}
\centering
\includegraphics[width=0.45\textwidth]{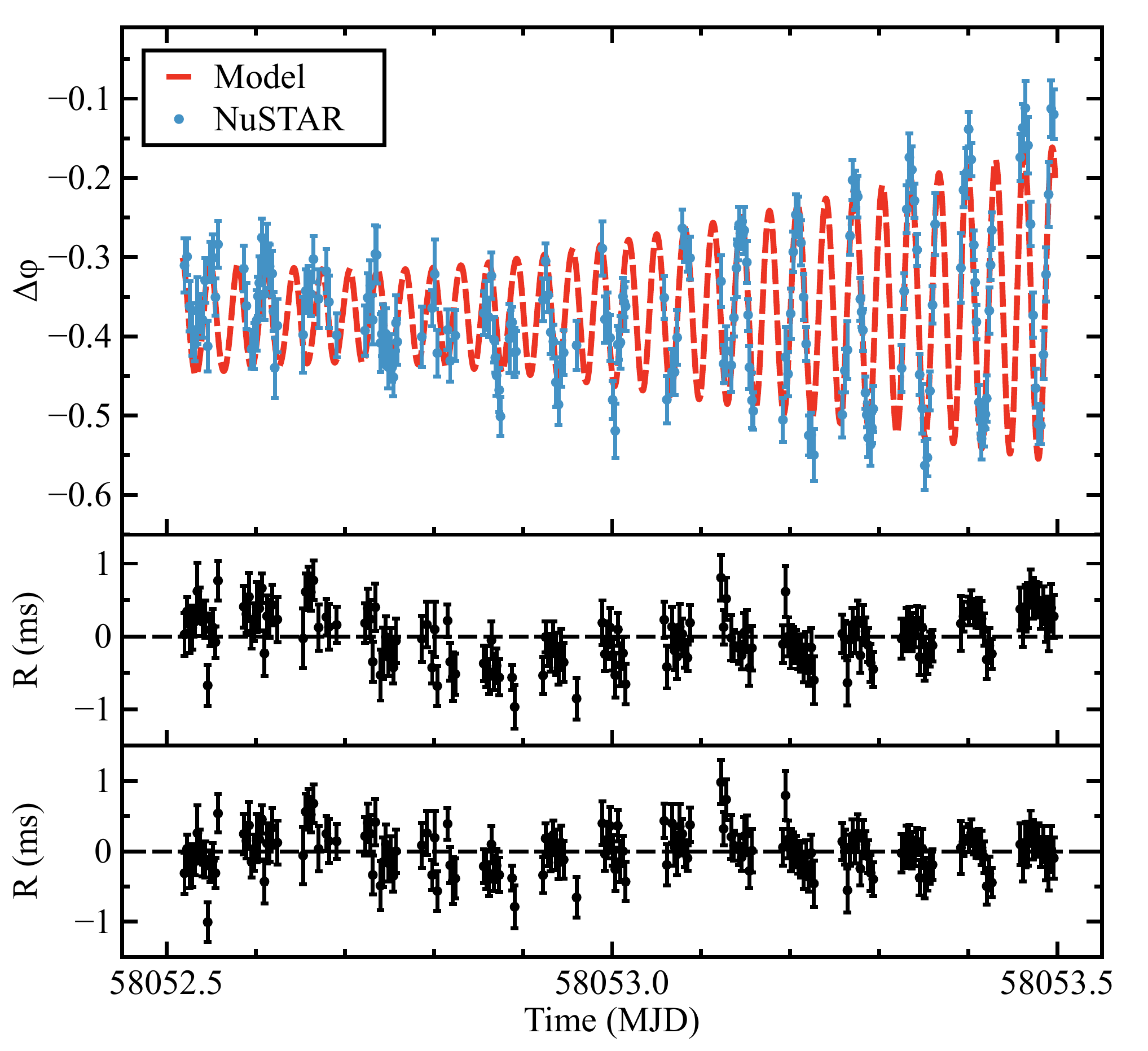}
\caption{\textit{Top panel -} Pulse phase delays as a function of time computed by epoch-folding $\sim$250 s-long intervals of  
\nustar{} data, together with the best-fit model (red dotted line, see text). \textit{Middle panel -} 
Residuals in ms with respect to the best-fit orbital solution including a linear model for the pulse phase delays. 
\textit{Bottom panel -} Residuals in ms with respect to the best-fit orbital solution including a quadratic model 
for the pulse phase delays.}
\label{fig:phase_fit}
\end{figure}

To investigate the temporal evolution of the pulse phase delays we used:
\begin{equation}
\label{eq:ph}
\Delta \phi(t)=\phi_0+\Delta \nu_0\,(t-T_0)-\frac{1}{2}\dot{\nu}\,(t-T_0)^2+R_{\rm orb}(t)
\end{equation}
where $T_0$ represents the reference epoch for the timing solution, $\Delta \nu_0=(\nu_0-\bar{\nu})$ is the correction at 
the reference epoch of the spin frequency used to epoch-fold the data, $\dot{\nu}$ is the spin frequency derivative, 
and $R_{\rm orb}$ is the Roemer delay generated by the differential corrections to the ephemeris applied to correct the photon 
time of arrivals \citep[e.g.][]{Deeter81}. The described process was iterated for each new set of orbital 
parameters obtained from the analysis, until no significant differential corrections were found for the parameters of 
the model. Best-fit parameters with and without a spin frequency derivative are shown in Tab.~\ref{tab:solution}. The measured derivative is several orders 
of magnitudes larger than typically observed in AMXPs, and we thus ascribe it to the \nustar{} internal clock 
drift, following a number of previous findings in the literature \citep{Madsen15, Sanna2017a, Sanna2017d, Sanna2017b}. 
Figure~\ref{fig:phase_fit} shows  
the pulse phase delays with the best-fitting model (top panel), and the corresponding residuals with respect to 
the model including a linear (middle panel) and a quadratic component (bottom panel) to fit the time evolution 
of pulse phase delays, respectively. 

\begin{figure}
  \includegraphics[width=0.49\textwidth]{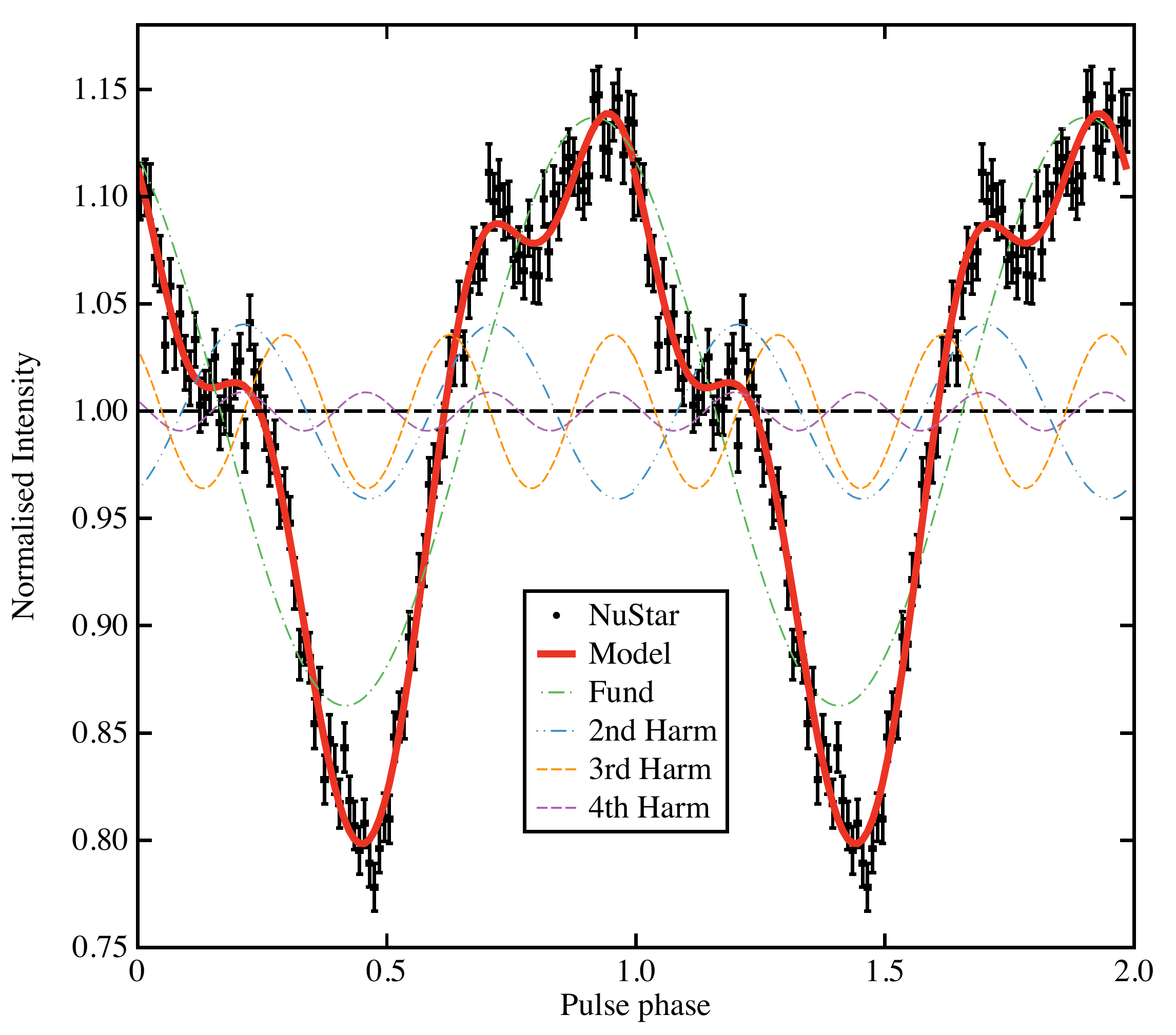}
  \caption{Pulse profile (black points) obtained from the epoch-folded \nustar{} data. The best fit model 
  obtained by combining four sinusoids with harmonically related periods is also shown (red line).  
Two cycles of the pulse profile are shown for clarity.}     
  \label{fig:profile}
\end{figure}

In Fig.~\ref{fig:profile} we show the best pulse profile obtained by epoch-folding the \nustar{} data with the parameters 
reported in Tab.~\ref{tab:solution} and sampling the signal in 100 phase bins. The peculiar pulse shape is well fitted with 
a combination of four sinusoids, where the fundamental, second, third and fourth harmonics have fractional amplitudes of 
$\sim14\%$, $\sim4\%$, $\sim3.8\%$ and $\sim0.9\%$, respectively. The presence of harmonics of the fundamental frequency is consistent 
with the results shown in the power density spectrum in Fig.~\ref{fig:pds}. 

Using the orbital binary parameters estimated from the \nustar{} timing analysis, we corrected the photon arrival times 
collected with \swiftxrt{}. We applied epoch-folding search techniques around the spin frequency reported in Tab.~\ref{tab:solution} 
with frequency steps of $10^{-5}$ Hz for a total of 101 steps. We significantly detected X-ray pulsations at the frequency 
$\nu=105.1757(2)$ Hz, consistent within errors with the value obtained from the \nustar{} data.

\begin{table}
\scriptsize
\centering
\caption{Orbital parameters and spin frequency of \igr{} estimated from the \nustar{} data. 
Left: without including a frequency derivative. Right: including a frequency derivative. Errors are at 
1$\sigma$ confidence level. Uncertainties are also scaled by a factor $\sqrt{\chi^2_{\rm red}}$ to take into account the large value of the reduced $\chi^2$.} 
\begin{tabular}{l | c c }
Parameters             & \multicolumn{2}{c} \nustar{} \\
\hline
\hline
R.A. (J2000) &  \multicolumn{2}{c}{$16^h59^m32.9007^s \pm 0.0021^s$}\\
DEC (J2000) & \multicolumn{2}{c}{$-37^\circ07\arcmin 14.22\arcsec \pm 0.18\arcsec$}\\
$P_{\rm orb}$ (s) &2758.3(3)&2758.2(3)\\
$x$ (lt-s) &0.00482(4)&0.00480(3)\\
$T_{\rm NOD}$ (MJD) & 58052.4892(2)&58052.4892(1)\\
e &$ < 4 \times 10^{-2}$&$< 4 \times 10^{-2}$\\
$\nu_0$ (Hz) &105.1758240(1)&105.1758271(3)\\
$\dot{\nu}$ (Hz/s) &--&-7.2(8)$\times 10^{-11}$\\
$T_0$ (MJD) & 58052.5 &58052.5\\
\hline
$\chi^2_\nu$/d.o.f. & 315.14/165 &210.37/164\\
\end{tabular}
\label{tab:solution}
\end{table}

\subsection{Spectral analysis}
\label{sec:spectral}
\begin{figure}
  \includegraphics[width=0.51\textwidth]{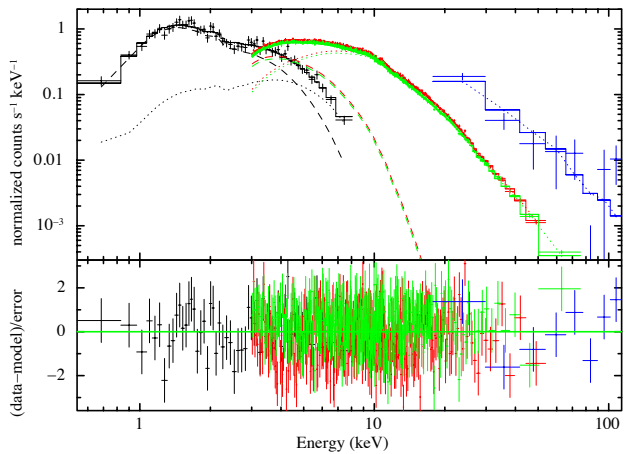}
  \caption{Broad-band spectrum of \igr\ (\swift{} in black, \nustar{} FPMA (FPMB) in red (green), and \ibis\ in blue). 
  The dashed and dotted lines represent the disk-blackbody and comptonization components of our model, respectively.}     
  \label{fig:spectrum}
\end{figure}

We performed spectral analysis with Xspec 12.9.1n \citep{Arnaud96} and fit near-simultaneous spectra in the 0.5-10 keV range for \swiftxrt{}, 
3-78 keV for \nustar{}, and 20-100 keV for \ibis{} (Fig.~\ref{fig:spectrum}). 
We assumed \citet{Wilms00} elemental abundances and \citet{Verner96} photo-electric cross-sections to model the interstellar medium.
We allowed for a normalisation coefficient between instruments, to account for cross-instrument calibration offsets and the  possibility of source variation. 

The spectra are well-described ($\chi^2_{\rm red}$/d.o.f.=1.02/652) by an absorbed disk blackbody plus thermally comptonized continuum with seed 
photons from the blackbody radiation ({\texttt const$\times$tbabs$\times$[diskbb+nthcomp]} in Xspec). The measured absorption column density 
of $(8.2\pm1.0)\times10^{21}$~cm$^{-2}$ is consistent with that expected in the direction of the source ($\sim 9.5\times10^{21}$ cm$^{-2}$) 
using the cluster A$_V$ \citep[][2010 revision]{Harris1996a} and the appropriate conversion 
from A$_V$ to N$_H$ assuming Wilms abundances \citep{Bahramian15, Foight2016a}. We obtained from the fit 
an inner disk temperature of $1.42\pm0.07$ keV, a power-law photon index of $2.3^{+0.2}_{-0.1}$, 
a blackbody seed photon temperature of $2.6\pm0.1$ keV, and a (poorly constrained) electron temperature of $\sim 30$ keV. 
We found no evidence for spectral lines (e.g., Iron K-$\alpha$) or reflection humps.

\section{Discussion}

With the discovery of pulsations at $\sim$105~Hz from \igr{} and the measurement of its orbital period at almost 46 min, 
we identify the source as the 20th known AMXP and a new member of the ultracompact low mass X-ray binaries. 
The X-ray spectrum of \igr{} is typical for an AMXP in outburst \citep[e.g.][]{Falanga2013a} and the non-detection 
of radio pulsations at the time of the Parkes observation is compatible with the fact that the X-ray outburst was still 
on-going. Assuming a typical pulsar spectral index of 1.6, the source previously detected at the VLA at 10 GHz \citep{Tetarenko2017a} would have a 1.4 GHz flux of 0.4 mJy. This would imply that the emission detected at the VLA is not pulsed and is most likely related to a radio-jet.

The system mass function $f(m_2, m_1, i)\sim1.2\times 10^{-7}$~M$_{\odot}$ and the lack of eclipses/dips in the 
X-ray light curve of the source allow us to constrain the mass of the donor star ($m_2$).  
Considering an upper limit on the system inclination angle of $i \lesssim 75^{\circ}$ (suggested by the absence of eclipses or dips in its light-curve) and assuming a 1.4~M$_{\odot}$ (2~M$_{\odot}$) NS, 
we obtain  $m_2 \gtrsim 0.0065$~M$_{\odot}$ ($m_2 \gtrsim 0.008$~M$_{\odot}$). 
This is consistent with the expected companion mass of $\sim$0.01 \Msun\ for an ultracompact NS binary in a 46-minute 
orbital period \citep[e.g.][]{van-Haaften2012a}. While the orbital properties of the system allow for a lower inclination angle and a 
higher mass, the requirement that the star also must fill its Roche lobe disfavours the higher mass solutions.

Given the measured unabsorbed flux of $\sim6.5\times 10^{-10}$ erg s$^{-1}$ cm$^2$ (0.5--100 keV), 
we estimate a source luminosity $L=6.5\times10^{36}$ erg s$^{-1}$. 
Under the assumption that the torques on the accreting X-ray pulsar are in equilibrium, we make a rough estimate of the dipolar 
magnetic field of the NS:
\begin{equation}
\label{eq:spineq}
B=4.2\,\zeta^{-7/4}\left(\frac{P_{\text{spin}}}{10 \,\text{ms}}\right)^{7/6}\left(\frac{M}{1.4M_{\odot}}\right)^{1/3}\left(\frac{\dot{M}}{10^{-10}M_{\odot} \text{yr}^{-1}}\right)^{1/2}10^8 \, \text{G},
\end{equation}
where $\zeta$ is a model-dependent dimensionless factor ranging between 0.1 and 1 that describes the relation between 
the magnetospheric radius and the Alfv\'en radius \citep[see e.g.,][]{Ghosh79a,Wang96,Bozzo2009a}, $P_{\text{spin}}$ represents 
the pulsar spin period in ms, $M$ is the NS mass and $\dot{M}$ is the rate of mass accreted onto the NS surface. 
Assuming a NS with $R=10$ km and mass $M=1.4$ M$_{\odot}$, we obtain  
$\dot{M}\simeq5.5\times10^{-10}$ M$_{\odot}$ yr$^{-1}$ and a dipolar magnetic field of $9.2\times10^8< B<5.2\times 10^{10}$ G. 
This is significantly larger than the average magnetic field of known AMXPs \citep[see e.g.,][]{Mukherjee2015a, Degenaar2017b}. Combined with the higher-than-average spin period of the source, 
this magnetic field suggests that \igr\ is being observed in a relatively early stage of its recycling process. 
The mass required to spin-up an old slowly-rotating neutron star up to $\sim105$ Hz \citep[of the 
order of $10^{-3}$ M$_{\odot}$; e.g.,][]{Burderi1999a} does not efficiently suppress the dipolar magnetic field, 
which limits the neutron star spin period. The large magnetic field combined with the moderately long spin period 
could be also responsible for the structured pulse profile, as well as the lack of emission lines and reflection components 
in the energy spectrum.

\begin{acknowledgements}
We thank the \nustar{}, \swift{}, and {\em Parkes} teams for the rapid scheduling of the ToO observation of \igr{}. DA acknowledges support from the Royal Society. ND is 
supported by a Vidi grant awarded by the Netherlands organization for scientific research (NWO). We acknowledge financial contribution from the agreement ASI-INAF I/037/12/0, and the International Space Science Institute (ISSI) Bern, which funded and hosted the international team ``The disk-magnetosphere interaction around transitional millisecond pulsars''. The Parkes radio telescope is part of the Australia Telescope National Facility which is funded by the Australian Government for operation as a National Facility managed by CSIRO. COH and GRS are supported by NSERC Discovery Grants. A. P. acknowledges funding from the EUs Horizon 2020 Framework Programme for Research and Innovation under the Marie Sk\l{}odowska-Curie Individual Fellowship grant agreement 660657-TMSP-H2020-MSCA-IF-2014.
\end{acknowledgements}

\bibliographystyle{aa} 
\bibliography{biblio.bib}

\end{document}